\documentclass[]{elsart}
\usepackage{graphicx}
\usepackage[dvips]{epsfig}
\usepackage{pstcol,pst-text}
\usepackage{latexsym}
\usepackage{amsfonts}
\usepackage{amsmath}
\newcommand{\bc}{\begin{center}}
\newcommand{\ec}{\end{center}}
\newcommand{\bq}{\begin{quote}}
\newcommand{\eq}{\end{quote}}
\newcommand{\bi}{\begin{itemize}}
\newcommand{\ei}{\end{itemize}}
\newcommand{\be}{\begin{equation}}
\newcommand{\ee}{\end{equation}}
\newcommand{\bea}{\begin{eqnarray}}
\newcommand{\eea}{\end{eqnarray}}
\newcommand{\bt}{\begin{tabular}}
\newcommand{\et}{\end{tabular}}
\newcommand{\btab}{\begin{tabbing}}
\newcommand{\etab}{\end{tabbing}}
\newcommand{\x}{\:\!}
\newcommand{\z}{\;\!}

\def\nn{\nonumber}
\def\mr{\mathrm}
\def\cal{\mathcal}

\begin{document}
\begin{flushright}
  {\small CERN--TH/2003--272}
\end{flushright}
\begin{frontmatter}
\title{A modified cluster-hadronization model\thanksref{mariecurie}}
\thanks[mariecurie]{Work supported in part by the EC 5th Framework
                    Programme under contract number HPMF-CT-2002-01663.}
\author[Dresden]{Jan-Christopher Winter},
\ead{\\winter@theory.phy.tu-dresden.de}
\author[Dresden,CERN]{Frank Krauss},
\ead{krauss@theory.phy.tu-dresden.de}
\author[Dresden]{Gerhard Soff}
\ead{Soff@physik.tu-dresden.de}
\address[Dresden]{Institut f{\"u}r Theoretische Physik,
                  TU Dresden,
                  D-01062 Dresden,
                  Germany}
\address[CERN]{Theory Division,
               CERN,
               CH-1211 Geneva 23,
               Switzerland}
\begin{abstract}
A new phenomenological cluster-hadronization model is presented. Its
specific features are the incorporation of soft colour reconnection, a
more general treatment of diquarks including their spin and giving
rise to clusters with baryonic quantum numbers, and a dynamic
separation of the regimes of clusters and hadrons according to their
masses and flavours. The distinction between the two regions
automatically leads to different cluster decay and transformation
modes. Additionally, these aspects require an extension of individual
cluster-decay channels that were available in previous versions of
such models.
\end{abstract}
\end{frontmatter}
%
%
%
%
\section{Introductory note}\label{sec_in}
Multi-hadron and jet production in high-energy particle reactions is a
basic property of the strong interaction
\cite{Dokshitzer:wu,Ellis:qj}. A successful description relies on a
factorization, which permits the separation of the perturbative
evolution from the non-perturbative development of an event. The
perturbative regime can be characterized through calculations of hard
matrix elements and subsequent multiple parton emissions -- the
physically appealing parton-shower picture%
\footnote{Perturbative QCD cascades can be formulated in two
          complementary ways, either in terms of quarks and gluons or
          in terms of colour dipoles \cite{cdmrefs,Lonnblad:1992tz}.}.
However the entire hadron-production mechanism cannot be precisely
predicted because of the lack of the understanding of non-perturbative
QCD effects, i.e. hadronization. For the transition of a coloured
partonic system into colourless primary hadrons, this implies a need
for phenomenological models. Lastly, after the primary-hadron genesis,
decays of unstable hadrons are accomplished. Employing the separation
ansatz, Monte Carlo event generators such as {\tt JETSET/PYTHIA}
\cite{Sjostrand:2000wi2001yu2003wg} or {\tt HERWIG}
\cite{Corcella:2000bw2002jc} proved to be a successful tool for the
description of multiparticle generation in high-energy physics.
\\
Concerning the transition process, such Monte Carlo schemes are either
based on the Feynman--Field or independent fragmentation
\cite{Field:fave}, on the Lund string \cite{Andersson:tv} and UCLA
\cite{Buchanan:1987ua:Chun:1992qs:Chun:bh} model ({\tt
JETSET/PYTHIA}), or on the cluster-hadronization model ({\tt HERWIG}).
The latter concept%
\footnote{Recent developments may be found, e.g. in
          \cite{recentdevs}.},
initially proposed by Wolfram and Field
\cite{Wolfram:1980gg,Field:1982dg}, and further advanced, among others
\cite{Gottschalk:1982ytfmbv}, by Webber and Marchesini
\cite{Webber:1983if,Marchesini:1983bm1987cf:Webber:1999ui}, explicitly
rests upon the preconfinement property of QCD
\cite{Amati:1979fg:Marchesini:1980cr} and the LPHD hypothesis
\cite{Azimov:1984np}. Such cluster models are usually formulated in
terms of two phases: cluster formation accomplished through the
non-perturbative splitting of gluons left by the parton shower into
quark--antiquark pairs, and cluster decays leading to the additional
creation of light-flavour pairs.
\\
To understand the physics at present and future colliders, e.g. the
Tevatron at Fermilab and the LHC at CERN, one fundamental cornerstone
is the implementation of new Monte Carlo event generators,
e.g. {\tt PYTHIA7} \cite{mc4lhc,Bertini:2000uh,Lonnblad:online}, and
{\tt HERWIG++} \cite{mc4lhc,Gieseke:2002sg,Gieseke:online}. The
development of the {\tt C++} Monte Carlo event generator {\tt SHERPA}
(Simulation of High Energy Reactions of PArticles)
\cite{mc4lhc,Krauss:online} is a step in the same direction. The
modified phenomenological cluster-hadronization model presented in
this paper contributes as a further module to the construction of the
{\tt SHERPA} package. The basic features of the new model are:
\\
Soft colour reconnection is accounted for in the formation and decay
of clusters. The flavour-dependent separation of the cluster regime
from the region of hadron resonances yields the selection of specific
cluster-transition modes. The two regimes are distinguished by
comparing the mass of the cluster with the masses of the accessible
hadrons matching the cluster's flavour structure.
\\
So far, the cluster scheme presented here is implemented only for
electron--positron annihilation, and, for simplicity, only the
light-quark sector is considered. An extension to heavy quarks,
however, is straightforward.
\\
The paper describing our cluster-hadronization model is organized as
follows:
first, different aspects of cluster formation are discussed in
Sec.~\ref{sec_cf}. Subsequently, in Sec.~\ref{sec_pa}, the
parametrization of light-flavour pair creation is presented. The
model's description is concluded by exhibiting cluster transformation
and fragmentation processes, which lead to the emergence of primary
hadrons, see Sec.~\ref{sec_cd}. The first results obtained with the
new hadronization scheme are shown in Sec.~\ref{sec_pr} for the
process $e^+e^-\!\to\gamma^{\star}/Z^0\!\to
d\bar{d},u\bar{u},s\bar{s}\to\mr{hadron\ jets}$.
\section{Cluster formation}\label{sec_cf}
The parton shower describes multiple parton emission in a
probabilistic fashion \cite{Ellis:qj}. By factorizing the full
radiation pattern into individual emissions it employs the
large-$N_{\mr{C}}$ limit of QCD. This organizes a binary tree, i.e. a
planar structure, of the partons. It also ensures that, once the
colour structure of the initial partons from the hard matrix element
is fixed, the colour structure of the partons at the end of the parton
shower is unambiguously determined.
\\
In our model, the non-perturbative transition of these partons into
primary hadronic matter, clusters, is accomplished by the following
steps:
\begin{enumerate}
\item 
To guarantee the independence of the hadronization model from the
quark masses eventually used in the parton shower and to account for a
gluon mass needed by the model, all partons are brought to their
constituent masses \cite{Webber:1983if}, $\cal{O}\x(0.3\ \mr{GeV})$,
$\cal{O}\x(0.3\ \mr{GeV})$ and $\cal{O}\x(0.45\ \mr{GeV})$ for $u$,
$d$\/ and $s$\/ flavours, and $\cal{O}\x(1\ \mr{GeV})$ for the gluon,
respectively. For this transition a numerical method, involving
several particles and consisting of a series of boosts and scaling
transformations, is employed. However, these manipulations are applied
only to parton-shower subsets that are in a colour-singlet state.
\item
Since in cluster-hadronization models the clusters consist of two
constituents in a colour-neutral state made up of a
triplet--antitriplet, the gluons from the parton shower must split (at
least) into quark--antiquark pairs \cite{Field:1982dg}. So, a
transition -- in principle non-perturbative -- transition $g\to
q\bar{q},\z \overline{D}D$ into a light quark--antiquark pair
$q\bar{q}$\/ or a light antidiquark--diquark pair $\overline{D}D$ (see
Sec.~\ref{sec_pa}) is enforced for each gluon. The respective flavour
composition of the gluon's decay products is obtained with the same
mechanism as used for cluster decays; see Sec.~\ref{sec_pa}. Quarks or
diquarks that cannot be produced owing to too high masses are
discarded. The kinematical distribution obeys axial symmetry; the
energy fraction $z$\/ of the quark (antidiquark) w.r.t. the gluon is
given by a density proportional to $z^2+(1-z)^2$, i.e. the gluon
splitting function%
\footnote{Obviously, for antidiquark--diquark pairs, this is a
          simplistic assumption, since it neglects, at least, the
          different spin structure of diquark production.}.
The limits on $z$\/ are fixed only after the flavour of the decay
products has been selected.
\item
In contrast to the Webber model of cluster fragmentation
\cite{Webber:online}, our model may also incorporate soft colour
reconnection%
\footnote{Other soft colour reconnection models are presented, e.g. in
          \cite{Sjostrand:1993hi,Webber:1997iw}.}
effects by eventually re-arranging the colours of the partons forming
the clusters. Starting with a simple cascade, Fig.~\ref{twoway}
schematically shows the two options to arrange two colour neutral
clusters out of four quarks or diquarks.
\begin{figure}[t!]
  \vspace{-5mm}
    \bc\includegraphics[height=48mm,angle=0.0]{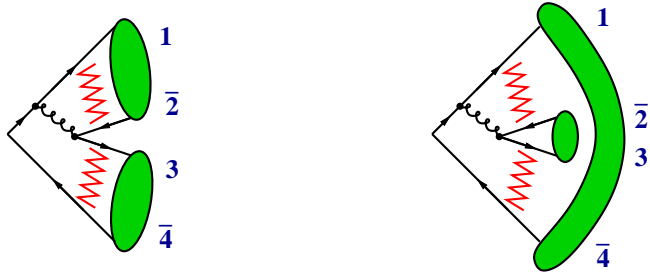}\ec
  \vspace{-5mm}
  \caption{Both options of cluster formation for a minimal
           $qg\bar{q}\to\!q\bar{q}'q'\bar{q}$\/ cascade. The zig-zag
           lines connecting the quark lines symbolize the soft
           exchange of colour quantum numbers, which is responsible
           for the colour reconnection.}
  \label{twoway}
  \vspace{0mm}
\end{figure}
The first -- direct -- case corresponds to the usual cluster formation
and reflects the leading term in the $1/N_{\mr{C}}$ expansion. The
second -- crossed -- configuration keeps track of subleading terms.
Motivated by the well-known colour suppression of non-planar diagrams
w.r.t. planar ones, the relative suppression factor due to colours is
taken to be $1/N^2_{\mr{C}}$. Additionally, a kinematical weight is
applied for each of the two possible cluster pairings. For the pairing
$ij,kl$ this weight reads
\be
  W_{ij,kl}\;=\;\frac{t_0}{t_0\,+\,4\z(w_{ij}+w_{kl})^2}\,,
  \label{KinWeight}
\ee
where the quantity $t_0$, of the order of $1\ \mr{GeV}^2$, denotes the
scale where the parton-shower evolution stops and hadronization sets
in. As a measure, $w_{ij}$ functions such as the invariant mass
\be
  m_{ij}\;=\;\sqrt{(p_i+p_j)^2\,}
\ee
of the parton pair (and therefore the cluster), or their relative
transverse momentum, similar to the Durham $k_{\perp}$ jet-scheme
\cite{Catani:1991hj}:
\be
  {p_{\perp}}_{ij}\;=\;\sqrt{\z2\z\min\{E^2_i,E^2_j\}\z(1-\cos\theta_{ij})\,}
\ee
might be used. The actual colour configuration of the considered
system is then chosen according to the combined colour and kinematical
weight. Ultimately, this reshuffling is iteratively applied to
combinations of two colour-singlet pairs of partons in the
colour-ordered chain.
\\
Of course, users who are not interested in colour reconnection have
the possibility to switch this option off.
\item
The cluster formation is accomplished by merging two colour-connected
partons, quark or antidiquark and antiquark or diquark, into a
colourless cluster. In this way, four different types may arise,
mesonic ($q_1\,\overline{q_2}$ and $\overline{D}\,D'$), baryonic
($q_1\,D$), and antibaryonic ($\overline{D}\,\overline{q_2}$)
clusters. The total four-momentum of these clusters is just given by
the sum of their constituent four-momenta \cite{Field:1982dg}.
\end{enumerate}
\section{Parametrization of light-flavour pair production}\label{sec_pa}
In our model the gluon splitting at the beginning of the cluster
formation phase and all cluster decays rely on the emergence of
light-flavour pairs, see \cite{Field:1982dg,Webber:1983if}. During
hadronization, which typically sets in at a scale of $1.0\ \mr{GeV}$,
there is no possibility for heavy-flavour pair generation
\cite{Andersson:tv}. The appearance of baryonic structures is tied to
the creation of light diquark--antidiquark pairs%
\footnote{Our treatment of diquarks resembles to some extent the one
          employed in the Lund approach for baryon production
          \cite{Andersson:1981ce:Andersson:1984af}.}.
In contrast to the Webber model, in our approach the total diquark
spin $S$\/ is explicitly considered. Thus, $q\bar{q}$\/ and
${\overline{D}}_SD_S$, where $q\in\{d,\,u,\,s\}$ and $D_S\in\{dd_1,\,
ud_0,\,ud_1,\,uu_1,\,sd_0,\,sd_1,\,su_0,\,su_1,\,ss_1\}$, occur as the
possible pairs. Apart from their masses influencing their emergence,
the created pair functions only as a flavour label. Furthermore, the
pair generation is assumed to factorize, i.e. to be independent of the
initial flavour configuration. Therefore, the only interest lies in
finding suitable pair-production probabilities, i.e. flavour and spin
symmetries should be correctly respected and reasonable hadron
multiplicities should be finally obtained in the hadron production.
\\
In our model a phenomenological parametrization is achieved by
employing hypotheses leading to a general ``flavour dicing'' scheme.
This scheme is applied to both regimes, cluster formation and decay.
The hypotheses are:
\begin{enumerate}
\item
The emergence of diquarks, i.e. baryons, is suppressed through a
factor $p_{\cal{B}}$ with $0\le p_{\cal{B}}\le1$.
\item
$\mr{SU}(3)_F$ symmetry is applied, but is assumed to be broken. This
is modelled by a strangeness suppression parameter $p_s$, $0\le
p_s\le1$, whereas the production of $u$\/ and $d$\/ flavours is
equally probable (strong-isospin symmetry), hence
$p_{u,d}=(1-p_s)/2$. As mentioned above, $p_{c,b}\equiv0$.
\item
Spin and flavour weights: the spin-$S$\/ diquark states ($S=0,1$) get
a weight proportional to $2\x S+1$. Additionally, a combinatorial
factor of $2$ and $1$ is applied, depending on whether different or
equal flavours constitute the diquark. But the fact that all states in
the baryonic $\mr{SU}(3)_F$ octet and decuplet appear equally likely
has to be reproduced. This gives rise to extra weights on the
individual diquark types. In particular, the combined diquark weights
$w^S_D$ read (up to the baryon suppression factor):
\be
  w^{S=0}_{D=ud,sd,su}\:=\;p_D\z,\quad\;\;
  w^{S=1}_{D=ud,sd,su}\:=\;3\,p_D\z,\quad\;\;
  w^{S=1}_{D=dd,uu,ss}\:=\;4\,p_D\z,
\ee
where
\be
  p_D\;=\;p_{d,u}^{2-n_s}\z\cdot\z p_s^{n_s}\z/\z(3\z p^2_s-2\z p_s+3)
\ee
and $n_s$ denotes the number of strange quarks in the diquark.
\\
An approach respecting $\mr{SU}(6)$ flavour-spin symmetry instead is
currently investigated.
\end{enumerate}
\section{Cluster transitions into primary hadrons}\label{sec_cd}
\begin{figure}[t!]
  \vspace{-5mm}
    \bc\includegraphics[height=42mm,angle=0.0]{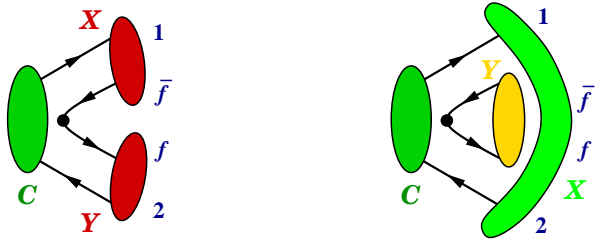}\ec
  \vspace{-5mm}
  \caption{Direct and crossed flavour arrangement and colour flow
           guaranteeing colour neutrality for each final-state
           configuration in cluster two-body decays.}
  \label{cluway}
  \vspace{0mm}
\end{figure}

Once the clusters have been formed, their masses are distributed
continuously and independently of the hard process with a peak at low
mass. In contrast, the observable hadrons have a discrete mass
spectrum and, hence, the clusters must be converted. This is achieved
through binary cluster decays and through transformations of
individual clusters into single primary hadrons. Our model does not
incorporate the subsequent decays of unstable hadrons. To model the
cluster transitions, the following assumptions are employed:
\begin{enumerate}
\item Cluster fragmentation is universal, i.e. independent of the hard
      process and of the parton shower. The clusters disintegrate
      locally without impact on other clusters.
\item Cluster transitions, i.e. decays as well as transformations,
      involve only low momentum transfer, of the order of $1\
      \mr{GeV}$ \cite{Webber:1983if}, since hadronization effects are
      supposed to be sufficiently soft and event-shape variables such
      as the thrust scale inversely with the centre-of-mass energy.
\item The regime of clusters is separated from the regime of hadrons
      according to the flavours of the cluster constituents and the
      accessible hadron masses. Clusters are supposed to be hadrons,
      if their mass is below a threshold mass. This bound is given by
      the maximum of the heaviest hadron with identical flavour
      content and the sum of the masses of the lightest possible
      hadron pair emerging in the decay of those clusters.
\end{enumerate}
The last assumption has two consequences, namely that in a first step
the newly formed clusters that are already in the hadronic regime have
to be transformed into hadrons; in the subsequent binary decays of the
remaining clusters, that also the daughter clusters, which fall into
the regime of hadron resonances, have to become hadrons immediately.
\\
In both cases, a definite hadron species $\cal H$\/ has to be chosen
according to the flavour structure of the considered cluster $\cal C$.
Respecting fixed particle properties, this choice is based on hadron
wave functions motivated by a non-relativistic quark model. The wave
functions are factorized into a flavour- and a spin-dependent part. In
our model the flavour part is given for a two-component system in
terms of quarks and diquarks. The overlap of this flavour part with
the flavour content of the cluster gives rise to a flavour weight. In
addition, since spin information is washed out in the clusters
\cite{Webber:1983if}, the total spin $J$\/ of the hadron manifests
itself as a corresponding weight. The total spin is given through the
coupling of the relative orbital momentum $L$\/ with the net spin
$S$\/ of the valence components. This can be written as $\vec J=\vec
L+\vec S$. The contributions of states with different orbital momentum
$L$\/ to the total-spin sum are accounted for by some a-priori weights
${\cal P}_L$, which enter as model parameters. Taken together, the
total flavour-spin weight ${\cal W}$\/ for a single hadron reads
\[
  {\cal W}\bigl(q_1s_1,\bar q_2s_2\to{\cal H}^J(q_1s_1,\bar q_2s_2)\bigr)
  \;\sim\hspace*{67mm}\nn\\[-3mm]
\]
\be
  \frac{\bigl|\langle\psi_F({\cal H}^J)|q_1s_1,\bar q_2s_2\rangle\bigr|^2}
       {\sum_{\hat{\cal H}|\hat J=J}
        \bigl|\langle\psi_F({\hat{\cal H}}^{\hat J})|
                     q_1s_1,\bar q_2s_2\rangle\bigr|^2}
  \,\cdot\,
  \frac{{\sum'}_{L,S\to J}
        \bigl|\langle S|s_1s_2\rangle\bigr|^2\,
        {\cal P}_L\,\bigl|\langle J|LS\rangle\bigr|^2}
       {\sum_{\hat J}{\sum'}_{\hat L,\hat S\to\hat J}
        \bigl|\langle\hat S|s_1s_2\rangle\bigr|^2\,
        {\cal P}_{\hat L}\,\bigl|\langle\hat J|\hat L\hat S\rangle\bigr|^2}\,.
  \label{totflavspinweight}
\ee
\\
In contrast to $q_1$ denoting the quarks, $\bar q_2$ stands for
antiquarks as well as diquarks. The spins of the two cluster
components $1$ and $2$ are given by $s_1=s(q_1)$ and $s_2=s(\bar
q_2)$, respectively, and $\langle\psi_F({\cal H}^J)|$ denotes the
flavour part of the hadron wave-function%
\footnote{For mesons this also includes the possibility of
          singlet-octet mixing occurring in hadron multiplets.}.
Moreover, $|\langle j|ls\rangle|^2=
2j+1/(\sum_{i=|l-s|,\ldots,(l+s)}2i+1)$ and ${\sum'}_{L,S\to J}$ is
an abbreviation denoting a summation over $L=0,1,\ldots$ and
$S=|s_1-s_2|,\ldots,(s_1+s_2)$, considering the condition that only
those terms contribute, where $|L-S|\le J\le(L+S)$ can be fulfilled.
Finally, it should be stressed that the second term of
Eq.~(\ref{totflavspinweight}) represents only a static model, which
accounts for the correct selection of hadrons according to their total
spin.
\\
The cluster fragmentation into primary hadrons is performed in two
phases:
\\
\hspace*{7mm}(I)\
When the clusters are formed from colour-connected pairs of quarks and
diquarks, some of them, because of their comparably low mass, fall
into the hadronic regime. Within our framework these clusters are
transformed into single hadrons immediately. In doing so, however,
some four-momentum is released and has to be absorbed by other
clusters. By allowing hadrons with masses lower than the cluster mass
only, the momentum transfer is taken to be a mere energy transfer and,
therefore, is time-like. This ensures that the absorbing cluster
becomes heavier. To fulfil the low momentum-transfer requirement, the
already outlined hadron-selection procedure according to the
flavour-spin weights ${\cal W}$\/ is modified through the inclusion of
an additional -- kinematic -- weight, which behaves like
\be
  {\cal W}_{\mr{kin.}}\;=\;
    \exp\left[-\left(\frac{Q^2}{Q^2_0}\right)^2\right]\,.
  \label{TransformWeight}
\ee
In this equation $Q^2>0$ denotes the squared momentum (i.e. energy)
transfer, and $Q_0$ is the scale related to the low momentum-transfer
demand. The limit $Q_0$, furthermore, depends on the cluster mass and
is also employed in the cluster decays; see below. Note that in the
Webber model the clusters being too light to decay are identified to
be the lightest hadron with identical flavour structure
\cite{Webber:online}. In comparison with the Webber scheme, the major
difference of our approach in the case of single-cluster transitions
is the expansion of the hadron-selection procedure.
\\
The cluster compensating the residual four-momentum is selected such
that it contains the partner that emerged in the same non-perturbative
gluon-splitting process as one of the constituents of the transformed
cluster. In turn, clusters, which fall into the hadron regime and
contain two leading quarks, are always split non-perturbatively into
two clusters containing only one leading constituent. In this context
leading partons, however, are only those quarks and antiquarks that
directly originate from the perturbative phase, and not from the
non-perturbative gluon splitting or from the cluster decays. For the
resulting single-leading clusters, then, the same considerations as
for the direct transformation to hadrons apply. In case a cluster in
the hadron regime is made of a diquark and an antidiquark, which is,
in principle, possible, it is forced to specifically decay into two
mesons. The details on the forced double-leading cluster breakup and
the double-diquark cluster decay are outlined below in paragraph (II);
see Eq.~(\ref{PrimKin}).
\\
\hspace*{7mm}(II)\
Finally all remaining primary and secondary (daughter) clusters have
to be split. The mass categorization outlined above automatically
yields one of the modes ${\cal C}\to{\cal C}_1{\cal C}_2$,
${\cal C}\to{\cal C}_1{\cal H}_2$, ${\cal C}\to{\cal H}_1{\cal C}_2$,
or ${\cal C}\to{\cal H}_1{\cal H}_2$. These modes involve the creation
of an extra flavour pair according to the ideas illustrated in
Sec.~\ref{sec_pa}. Similarly to the cluster-formation phase, then, two
flavour configurations for the decay products emerge, namely a direct
one and a crossed one; see Fig.~\ref{cluway}. Again, the crossed
configuration is suppressed by the colour factor $1/N^2_{\mr C}$ and
the kinematical weight from Eq.~(\ref{KinWeight}) using identical
measure functions $w$\/ and replacing $t_0$ by $Q^2_0$, which again
depends on the mass of the decaying cluster.
The cluster-decay kinematics, which makes use of the parameter $Q_0$,
is fixed to be anisotropic. Starting from a mother cluster with
constituent momenta $p^{\cal C}_{1,2}$, the new momenta of the
decay-products' constituents read \cite{Webber:1983if}
\be
  p_{1,2}\;=\;\left(1-\frac{Q_0}{M_{\cal C}}\right)\,p^{\cal C}_{1,2}\,,\qquad
  p_{\bar f,f}\;=\;\frac{Q_0}{M_{\cal C}}\,p^{\cal C}_{2,1}\,,
  \label{PrimKin}
\ee
where $f$\/ and $\bar f$\/ label the momenta of the newly created
flavour pair. Hence, for the two cluster arrangements (see
Fig.~\ref{cluway}) the momenta are given by $P^{\cal
X}_{\mr{dir.}}=p_1+p_{\bar f},\,P^{\cal Y}_{\mr{dir.}}=p_f+p_2$ in the
direct, and $P^{\cal X}_{\mr{cross.}}=p_1+p_2,\,P^{\cal
Y}_{\mr{cross.}}=p_f+p_{\bar f}$ in the crossed case, respectively. To
guarantee well-behaved four-momenta in this fission breaking, our
model uses a running $Q_0$ depending on two parameters, $\hat Q_0$ and
$\hat M_0$, with the constraint $\hat Q_0<\hat M_0$:
\be
  Q_0(M_{\cal C})\;=\;\frac{\hat Q_0\cdot M_C}{\hat M_0+M_{\cal C}}
                 \;<\;M_{\cal C}\,.
\ee
Having fixed the primary kinematics, via Eq.~(\ref{PrimKin}), and the
combination of flavours and momenta to the new clusters, their masses
can be deduced from the squares of their total four-momenta. Then, as
stated above, the different decay modes ${\cal C}\to
{\cal C}_1{\cal C}_2,\,{\cal C}_1{\cal H}_2,\,
{\cal H}_1{\cal C}_2,\,{\cal H}_1{\cal H}_2$ are distinguished. All
possible decay channels within each mode are comprehensively
summarized in Table~\ref{DecTypes}.
\begin{table}[t!]
  \caption{Different cluster types emerging through cluster breakups.
           Decay channels indicating a four-quark, i.e. two-diquark,
           system to become a hadron within the modes ${\cal C}\to
           {\cal C}_1{\cal H}_2,\,{\cal H}_1{\cal C}_2,\,
           {\cal H}_1{\cal H}_2$ are vetoed. The four-quark cluster
           disintegration into two mesons (see the last row of the
           table) is only available for the mode ${\cal C}\to
           {\cal H}_1{\cal H}_2$. The occurrence of the two
           disintegration possibilities is taken to be equally
           likely.}
  \label{DecTypes}
  \bc
    \begin{tabular}{|c|cll|cll|}\hline
    Cluster && Direct case & Crossed case && Direct case & Crossed case\\\hline
    $q_1\bar q_2$      & $\stackrel{\bar q q}{\longrightarrow}$ &
                         $q_1\bar q+q\bar q_2$, &
                         $q_1\bar q_2+q\bar q$ 
                       & $\stackrel{D\overline D}{\longrightarrow}$ &
                         $q_1 D+\overline D\bar q_2$, &
		         $q_1\bar q_2+\overline DD$\\
    $q_1 D_2$          & $\stackrel{\bar q q}{\longrightarrow}$ &
                         $q_1\bar q+qD_2$, &
                         $q_1D_2+q\bar q$
                       & $\stackrel{D\overline D}{\longrightarrow}$ &
                         $q_1 D+\overline DD_2$, &
                         $q_1D_2+\overline DD$\\
    $\overline D_1D_2$ & $\stackrel{\bar q q}{\longrightarrow}$ &
                         $\overline D_1\bar q+qD_2$, &
                         $\overline D_1D_2+q\bar q$
                       & $\stackrel{D\overline D}{\longrightarrow}$ &
                         $\overline D_1 D+\overline DD_2$, &
                         $\overline D_1D_2+\overline DD$\\
    $\overline D_1D_2$ & $\longrightarrow$ &
                         $q_2\bar q_1+q_2'\bar q_1'$ &
                       & $\longrightarrow$ &
                         $q_2\bar q_1'+q_2'\bar q_1$ &\\\hline
    \end{tabular}
  \ec
  \vspace{1.4mm}
\end{table}
\begin{enumerate}
\item For the case of breakups involving clusters only, i.e. for
      ${\cal C}\to{\cal C}_1{\cal C}_2$, nothing has to be done in
      addition.
\item If one of the daughter clusters falls into the hadronic regime,
      i.e. for ${\cal C}\to{\cal C}_1{\cal H}_2$ and ${\cal C}\to{\cal
      H}_1{\cal C}_2$, a suitable hadron has to be selected such that
      the hadron will be lighter than the cluster. The selection
      procedure follows the one outlined above for the ${\cal
      C}\to{\cal H}$ transformation; the recoil is taken by the
      daughter system, which belongs to the cluster regime.
\item If both new clusters fall into the hadron regime, i.e. for
      purely hadronic decays ${\cal C}\to{\cal H}_1{\cal H}_2$, more
      severe manipulations are applied. First of all, the newly
      created flavour pair $f\bar f$\/ is abandoned; instead, two
      hadrons are chosen directly. Then the combined weight for the
      selection of such a hadron pair consists of three pieces. The
      first part accounts for the two flavour-spin contents. The
      second one includes the correct relation of direct to crossed
      decay configurations and, furthermore, represents the
      incorporation of the pair-production rates. The last part
      considers the phase space of the decay, which is taken to be
      isotropic in the cluster's rest frame
      \cite{Field:1982dg,Webber:1983if}. The combination of the first
      two weights for the hadron pair is set up as if only complete
      $\mr{SU}(3)_F$ multiplets were accessible. Because of the
      superposition with the phase-space factor, a hadron pair that
      cannot be produced in a cluster decay owing to its large mass
      cannot contribute to the selection%
      \footnote{The weight treatment for hadron selection in {\tt
                HERWIG} was modified by Kup\v{c}o \cite{Kupco:1998fx}.
                However, currently, the {\tt HERWIG++} group is
                working on a new approach \cite{mc4lhc}.}.
      The other manipulation, as indicated above, is that once the
      hadron species are chosen, the cluster decays isotropically in
      its rest frame into these hadrons.
      \\
      Two comments are in order here: first of all, our approach takes
      leading-particle effects into account in the same manner as in
      Webber's model \cite{Webber:online}. Secondly, when considering
      a cluster consisting of two diquarks, mesons can emerge only by
      recombining the individual quarks and antiquarks that constitute
      the diquarks, see Table~\ref{DecTypes}. Since baryons appear in
      a decay of such clusters through the creation of a quark pair,
      the diquark recombination is taken to be suppressed by a factor
      of $p_{\cal B}$ w.r.t. the baryon production, which appears with
      $1-p_{\cal B}$ in this channel. The specific ordering of the
      quarks into mesons is then done in a fashion similar to the one
      above, involving hadron pairs. The difference, however, lies in
      the fact, that Clebsch-Gordan coefficients are additionally
      employed. These coefficients account for the rearrangement of
      the diquark spin wave-functions into a double-mesonic basis.
\end{enumerate} 
\section{Preliminary results}\label{sec_pr}
\begin{figure}[t!]
  \vspace{-36mm}
    \bc
      \includegraphics[height=120mm,width=102mm]{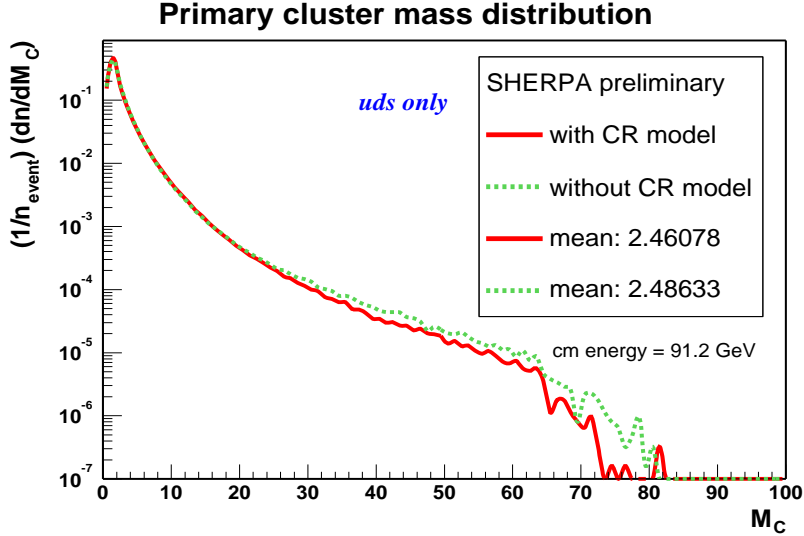}
    \ec
  \vspace{-13mm}
  \caption{Primary cluster-mass distribution in electron--positron
           annihilation events that evolve into light-quark and gluon
           jets at the $Z^0$ pole. The {\tt SHERPA$\alpha$} result is
           shown with (solid line) and without (dashed line)
           colour-reconnection (CR) model.}
  \label{massdist}
  \vspace{-1mm}
\end{figure}

\begin{figure}[t!]
  \vspace{-2mm}
    \bc\includegraphics[height=74mm,width=122mm]{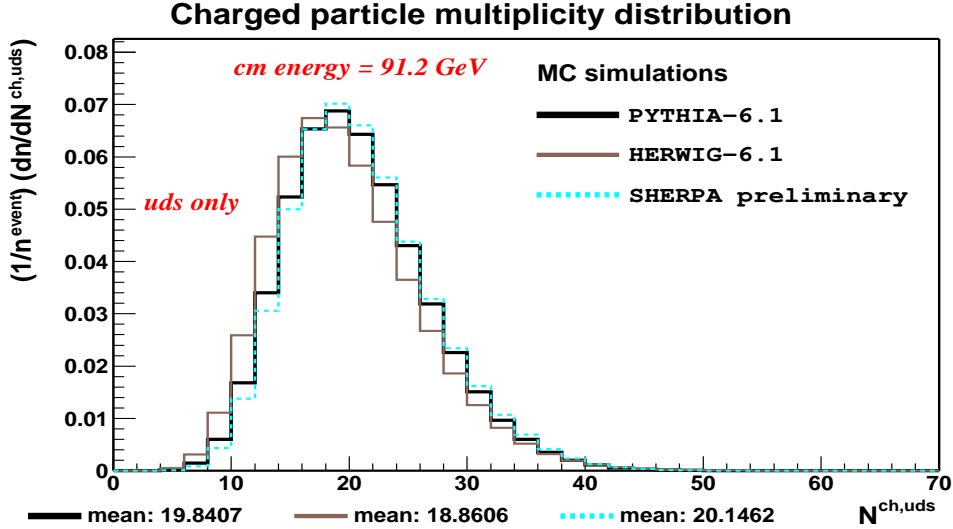}\ec
  \vspace{-1mm}
  \caption{Predicted multiplicity distribution of charged particles in
           $e^+e^-$ annihilation for light-quark and gluon jets at the
           $Z^0$ pole. The {\tt SHERPA$\alpha$} result is compared
           with the default {\tt PYTHIA-6.1($uds$)} and {\tt
           HERWIG-6.1($uds$)} predictions.}
  \label{multdist}
  \vspace{-1mm}
\end{figure}

\begin{figure}[t!]
  \vspace{-2mm}
    \bc\includegraphics[height=84mm]{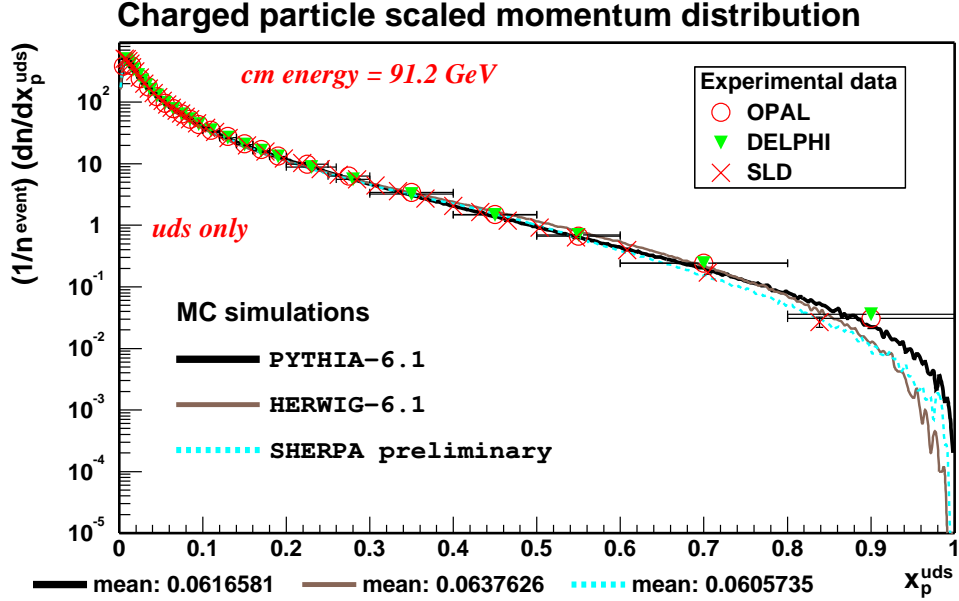}\ec
  \vspace{-1mm}
  \caption{Scaled momentum distribution of charged particles for
           $E_{\mr{cm}}=91.2\ \mr{GeV}$ in $e^+e^-$ annihilation
           considering only the light-quark sector. The {\tt
           SHERPA$\alpha$} prediction is compared with experimental
           light-quark data provided by the OPAL, DELPHI and SLD
           collaborations, and to the {\tt PYTHIA-6.1($uds$)} and {\tt
           HERWIG-6.1($uds$)} outcomes, using their default settings.
           Concerning the mean value $\langle x^{uds}_p\rangle$ of the
           distributions, only the {\tt HERWIG-6.1($uds$)} prediction
           is consistent with the OPAL measurement of $\langle
           x^{uds}_p\rangle=0.0630\pm0.0003(\mr{stat.})\pm0.0011
           (\mr{syst.})$ given in \cite{Ackerstaff:1998hz}.}
  \label{xpdist}
  \vspace{-1mm}
\end{figure}

\begin{figure}[t!]
  \vspace{-2mm}
    \bc\includegraphics[height=71mm,width=114mm]{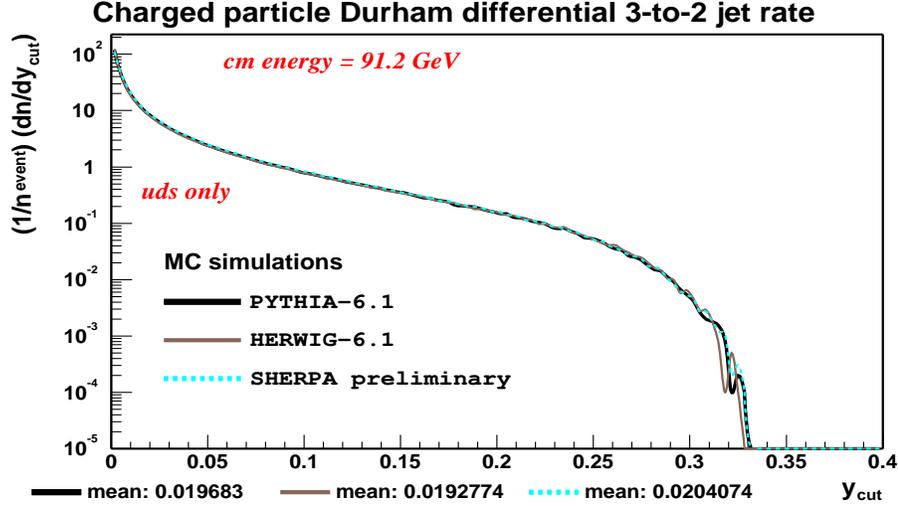}\ec
  \vspace{-1mm}
  \caption{The Durham $3\to2$ differential jet rate of charged
           particles in $e^+e^-$ annihilation at the $Z^0$ pole. Only
           $uds$\/ events are taken into account. The {\tt
           SHERPA$\alpha$} result is compared with the results
           stemming from {\tt PYTHIA-6.1($uds$)} and {\tt
           HERWIG-6.1($uds$)} performances, both of which run with
           their default parameters.}
  \label{durhamdist}
  \vspace{-1mm}
\end{figure}

The performance of the model introduced above is now illustrated by
presenting some results for $e^+e^-$ annihilation at the $Z^0$ pole
using only light quarks throughout the event's evolution. The outcomes
have been obtained with the parton shower of {\tt APACIC++-1.0}
\cite{Kuhn:2000dk}, the matrix elements are generated by {\tt
AMEGIC++-1.0} \cite{Krauss:2001iv} and matched with this parton shower
\cite{Catani:2001cc}, the primary hadronization is accomplished by the
cluster model described above, and the hadron decays are provided
through interfacing the corresponding routines of {\tt PYTHIA-6.1}
\cite{Sjostrand:2000wi}. The resulting event generator is the
combination of these modules. In the following it is referred to as
{\tt SHERPA$\alpha$}. All results shown below are achieved with the
same parameter set, where the cluster-model parameters have been
adjusted by hand. The settings of the other module's input variables
are mainly taken over from a tuning of {\tt APACIC++-1.0}, together
with the full hadronization of {\tt PYTHIA-6.1}. Since measurements
that specifically concentrate on the observation of light-quark
characteristics are rarely available, our results are mainly compared
with those gained from running {\tt PYTHIA-6.1} and {\tt HERWIG-6.1}
\cite{Corcella:1999qn} both restricted to $u,d,s$\/ quarks. Thereby
either of the models has been run with its default parameter values.
\\
To begin with, the effects of our colour-reconnection model on the
cluster-mass distribution, and the statistics of the reconnections in
the cluster formation are briefly discussed. Figure~\ref{massdist}
illustrates the statement that under the influence of colour
reconnection our cluster hadronization tends to produce less massive
primary clusters than without the reconnection procedure. The decrease
is especially caused by the kinematical factor, Eq.~(\ref{KinWeight}),
where $w_{ij}={p_{\perp}}_{ij}$ has been used. In the cluster
formation one gets approximately $0.74$ reconnections per event and,
with a frequency of $48\%,35\%,13\%,3\%$, and $1\%$, one finds
$0,1,2,3$, and $>3$ exchange(s), respectively. Moreover, switching off
the colour-reconnection option entirely while keeping all the other
parameters unchanged yields the following qualitative modifications:
the number of daughter clusters per event is increased, which results
in an enlargement of the mean charged-particle multiplicity of roughly
$0.2$ tracks per event. The charged-pion production rate increases
whereas the charged-kaon rate and the (anti)proton rate decrease.
Furthermore, the charged particle transverse-momentum distributions
are lowered for high $p^{\mr{in/out}}_{\perp}$; however, for the
scaled-momentum distribution of charged tracks, see also below, the
bump at $x^{uds}_p\approx0.5$ enhances and its tail tends to become
harder.
\\
The overall charged-particle multiplicity distribution is presented in
Fig.~\ref{multdist}. The shift to higher multiplicities of the {\tt
SHERPA$\alpha$} curve w.r.t. the other curves indicates the higher mean
value of the {\tt SHERPA$\alpha$} prediction. Table~\ref{multnum}
shows mean multiplicities $\langle\cal{N}^{uds}_{\mr{ch}}\rangle$ as
provided by those three fragmentation models in comparison with
inclusive measurements.
To exemplify the charged hadron rates, the mean multiplicities for the
eventually observable charged hadrons -- $\pi^{\pm}$, $\it{K}^{\pm}$
and $\it{p},\bar{\it{p}}$ -- are considered and compared with
experimental $uds$\/ results; see also Table~\ref{multnum}. In view of
these comparisons, one can conclude that the obtained {\tt
SHERPA$\alpha$} multiplicity results are satisfactory, and in good
agreement with the {\tt PYTHIA-6.1($uds$)} predictions as well as with
the data.
\begin{table}[b!]
  \caption{Overall mean charged-particle multiplicity, and production
           rates of charged pions, charged kaons and (anti)protons in
           $e^+e^-$ collisions. The values are taken for $uds$\/
           events running at the $Z^0$-peak centre-of-mass energy.
	   The errors indicated in the table are the total errors of
           the measurements. The abbreviations {\tt PY61($uds$)} and
           {\tt HW61($uds$)} stand for {\tt PYTHIA-6.1($uds$)} and
           {\tt HERWIG-6.1($uds$)}, respectively. More {\tt JETSET}
           and {\tt HERWIG} results on the topic can be found in
           \cite{Abreu:1998vq}.}
  \label{multnum}
  \bc
    \begin{tabular}{|c||c||c|c|c|}\hline    
      & $\langle\cal{N}^{uds}_{\mr{ch}}\rangle$
      & $\langle\cal{N}^{uds}_{\pi^{\pm}}\rangle$
      & $\langle\cal{N}^{uds}_{\it{K}^{\pm}}\rangle$
      & $\langle\cal{N}^{uds}_{\it{p},\bar{\it p}}\rangle$\\\hline
      \multicolumn{5}{c}{}\\[-6mm]\hline
      {\tt PY61($uds$)} &
      $19.84$ & $16.72$ & $2.010$ & $0.856$\\\hline
      {\tt HW61($uds$)} &
      $18.86$ & $15.37$ & $1.693$ & $1.568$\\\hline
      {\tt SHERPA$\alpha$} &
      $20.15$ & $16.83$ & $2.018$ & $1.047$\\\hline
      \multicolumn{5}{c}{}\\[-6mm]\hline
      OPAL   \cite{Ackerstaff:1998hz} &
      $20.25\pm0.39$ &&&\\\hline    
      DELPHI \cite{Abreu:1997ir} &
      $20.35\pm0.19$ &&&\\\hline    
      DELPHI \cite{Abreu:1998vq} &
      $19.94\pm0.34$ & $16.84\pm0.87$ & $2.02\pm0.07$ & $1.07\pm0.05$\\\hline
      SLD    \cite{Abe:1996zi} &
      $20.21\pm0.24$ &&&\\\hline    
      SLD    \cite{Abe:2003iy} &
      $20.048\pm0.316$ & $16.579\pm0.304$ &
      $2.000\pm0.068$ & $1.094\pm0.043$\\\hline
    \end{tabular}
  \ec
\end{table}
\\
As an example for a particle-momentum distribution the scaled momentum
$x^{uds}_p=2|\vec p_{uds}|/E_{\mr{cm}}$ and its negative logarithm
$\xi^{uds}_p=-\ln x^{uds}_p$ are considered. The $x^{uds}_p$
distribution obtained with {\tt SHERPA$\alpha$} is shown in
Fig.~\ref{xpdist}, and compared with the predictions of the {\tt
PYTHIA-6.1($uds$)} and {\tt HERWIG-6.1($uds$)} event generators.
Furthermore, experimental results delivered by the OPAL, DELPHI and
SLD collaborations on this differential cross section are included.
The {\tt PYTHIA-6.1($uds$)} model is the most consistent with the OPAL
and DELPHI data, but it predicts a slightly softer spectrum. Both
cluster-hadronization models show a similar behaviour concerning their
deviation from these data. For $x^{uds}_p<0.7$ they wiggle around the
{\tt PYTHIA-6.1($uds$)} prediction and for $x^{uds}_p>0.8$ they
anticipate a steeper decline, which is quite different from that seen
in the OPAL and DELPHI data. However, recently published SLD results
on this topic support the tendency of having a much weaker
high-$x^{uds}_p$ tail. This behaviour then is well described by our
cluster model. The hump at $x^{uds}_p\approx0.5$ is truly a deficiency
of cluster approaches. In comparison with the {\tt HERWIG-6.1($uds$)}
prediction, our model yields a smaller bump, and the values for
$x^{uds}_p>0.9$ do not fall off as rapidly as the {\tt
HERWIG-6.1($uds$)} ones. This performance might be due to the mass
categorization treatment of the cluster transitions, which has been
introduced in our model. All in all, the $x^{uds}_p$ behaviour clearly
reflects two cluster-model weaknesses, namely (1) that the necessary
increase in cluster and, therefore, in hadron multiplicity excessively
results in a decrease of large three-momenta of primary clusters, and
(2) that the hadronization of events with a small number of primary
clusters is not sufficiently modelled yet.
\begin{figure}[t!]
  \vspace{-2mm}
    \bc\includegraphics[height=84mm]{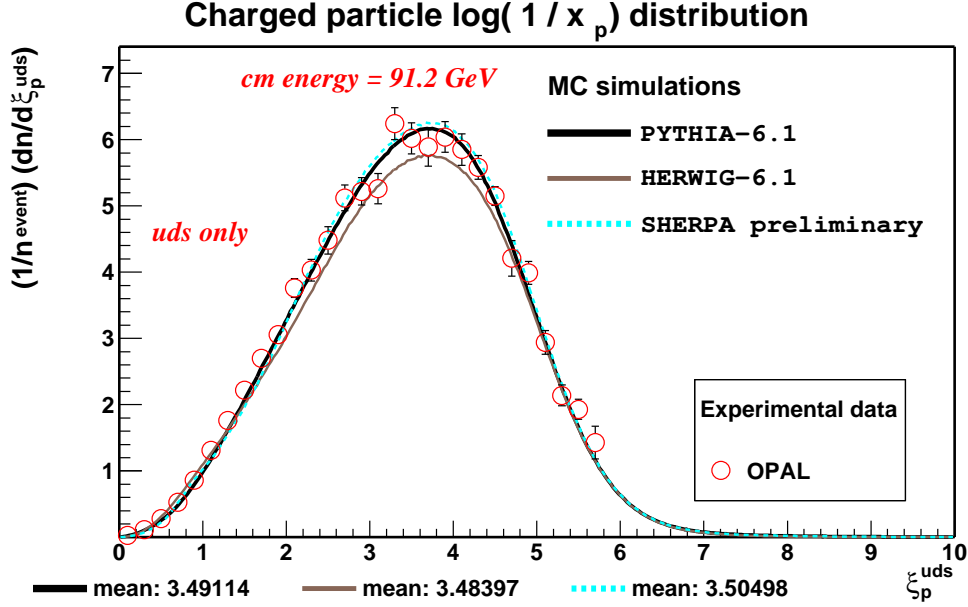}\ec
  \vspace{-1mm}
  \caption{$\xi^{uds}_p=\ln(1/x^{uds}_p)$ distribution of charged
           particles for $E_{\mr{cm}}=91.2\ \mr{GeV}$ in $e^+e^-$
           annihilation, considering the light-quark sector only. The
           {\tt SHERPA$\alpha$} prediction is presented together with
           experimental $uds$\/ data provided by the OPAL
           collaboration, and with results from default {\tt
           PYTHIA-6.1($uds$)} and default {\tt HERWIG-6.1($uds$)}.}
  \label{xipdist}
  \vspace{-1mm}
\end{figure}
\\
In contrast to the $x^{uds}_p$ distribution, the $\xi^{uds}_p$
distribution emphasizes the soft momenta of the spectrum.
Figure~\ref{xipdist} illuminates the {\tt SHERPA$\alpha$} result
together with those of the other two QCD Monte Carlo models, and
compares them with experimental measurements from the OPAL, DELPHI and
SLD collaborations. {\tt SHERPA$\alpha$} describes the data over most
of the $\xi^{uds}_p$ region, and is in quite good agreement with the
{\tt PYTHIA-6.1($uds$)} prediction. It underestimates the region of
$1<\xi^{uds}_p<2$; on the other hand, however, it slightly
overestimates the data for $3<\xi^{uds}_p<5$. Experimental inclusive
measurements of the peak position, $\xi^{\ast,uds}_p=3.76\pm0.02$
(DELPHI \cite{Abreu:1998vq}) and $\xi^{\ast,uds}_p=3.74\pm0.22$ (OPAL
\cite{Ackerstaff:1998hz}), seem to be reproduced by the {\tt
PYTHIA-6.1($uds$)} and {\tt SHERPA$\alpha$} Monte Carlo simulations.
{\tt HERWIG-6.1($uds$)} is considerably low (high) for
$2<\xi^{uds}_p<5$ ($0.5<\xi^{uds}_p<1$).
\begin{figure}[b!]
  \vspace{-2mm}
    \bc\includegraphics[height=71mm,width=114mm]{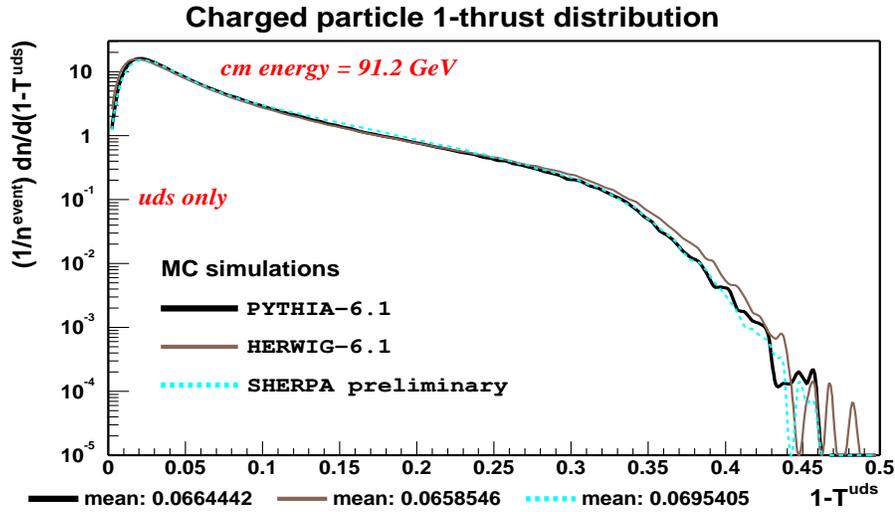}\ec
  \vspace{-1mm}
  \caption{$1-T/x^{uds}_p$ distribution of charged particles for
           $E_{\mr{cm}}=91.2\ \mr{GeV}$ in $e^+e^-$ annihilation with
           a restriction on $u,d,s$\/ and gluon jets. The {\tt
           SHERPA$\alpha$} prediction is compared with predictions of
           default {\tt PYTHIA-6.1($uds$)} and {\tt
           HERWIG-6.1($uds$)}.}
  \label{1-thrustdist}
  \vspace{-1mm}
\end{figure}
\\
As an example for the group of event-shape observables, the $1-T$\/
distribution, $T$\/ being the thrust, of the three aforementioned QCD
Monte Carlo event generators with $u,d,s$\/ quark restriction is
presented in Fig.~\ref{1-thrustdist} for light-quark and gluon jets.
{\tt HERWIG-6.1($uds$)} accounts on average for more spherical event
shapes, which is indicated by a weaker decline of the spectrum towards
higher values. Owing to the LPHD concept, the {\tt SHERPA$\alpha$}
prediction, somewhat exceeding the {\tt PYTHIA-6.1($uds$)} result for
$0.1<1-T<0.3$, rather resembles the prediction of {\tt
PYTHIA-6.1($uds$)}, which might be due to the fact that {\tt
SHERPA$\alpha$} employs a {\tt PYTHIA}-like parton shower.
\\
Lastly the Durham $3\to2$ differential jet rate is considered in
Fig.~\ref{durhamdist}. Except for the low-statistics region, the
results for the event generators shown in the plot barely exhibit any
deviation from one another.
\\
Taken together, one can conclude that a good performance could be
achieved by the new model in electron--positron annihilation into
light-quark and gluon jets at the $Z^0$ pole, although this model has
not been sufficiently tuned yet. These first {\tt SHERPA$\alpha$}
outcomes indicate an encouraging agreement with results obtained from
{\tt PYTHIA-6.1} restricted onto the light-quark sector. Where
provided, the comparison with experimental data is satisfactory.
\section{Summary and conclusions}
A modified cluster-hadronization model has been presented. In
comparison with the long-standing Webber model, the extensions of our
approach are the following.
\\
Soft colour-reconnection effects are included in the cluster formation
as well as in the cluster-decay processes. This yields an enhancement
of the number of decay configurations. The spin of diquarks is
explicitly accounted for throughout the model. The number of basic
cluster species is enlarged, especially by a new mesonic-cluster type,
the four-quark cluster. The significant feature of our approach is the
flavour-dependent separation of the cluster and hadron regimes in
terms of the mother cluster's mass. This categorization automatically
selects the cluster-transition mode. Taken together, these aspects
require the set-up of generically new cluster decay channels.
\\
Our cluster-hadronization model is implemented as a {\tt C++} code.
The resulting version is capable of describing electron--positron
annihilation $e^+e^-\!\to\gamma^{\star}/Z^0\!\to d\bar d,u\bar u,s
\bar s$\/ into light-quark and gluon jets. Some first tests were
passed (see previous section) and the agreement with {\tt
PYTHIA-6.1($uds$)} and experimental data is satisfactory. Some
cluster-model shortcomings, such as the too low charged-particle
multiplicity, could be cured; and the spectrum of the scaled momentum
could be improved. The model will soon be completed by including
heavy-quark hadronization. Furthermore, the focus of future work is on
treating the fragmentation of remnants of incoming hadrons, especially
in view of proton--(anti)proton applications (Tevatron and LHC
physics).
%
%
%
%
\begin{ack}
J.W. and F.K. would like to thank Bryan Webber, Torbj{\"o}rn
Sj{\"o}strand, Rick Field, Leif L{\"o}nnblad, Stefan Gieseke, Philip
Stevens, Alberto Ribon and Mike Seymour for fruitful and pleasant
communication on the subject.
\\
J.W. and F.K. are indebted to Klaus Hamacher and Hendrik Hoeth for
valuable discussions, especially on the outcome of the model.
\\
Special thanks go to Suzy Vascotto for carefully reading the
manuscript.
\\
J.W. would like to thank the TH division at CERN for kind hospitality
during the MC4LHC workshop, where parts of this work were completed.
\\
F.K. acknowledges financial support from the EC 5th Framework
Programme under contract number HPMF-CT-2002-01663. The authors are
grateful for additional financial support by GSI, BMBF and DFG.
\end{ack}

\end{document}